\documentclass[]{aa}

\def\rmit#1{{\it #1}}              
\def\specchar#1{{\sc #1}}
\def\FeI{\mbox{Fe\,\specchar{i}}}

\def\BaII{\mbox{Ba\,\specchar{ii}}}

\def\ie{\rmit{i.e.}}

\usepackage{natbib}
\usepackage[dvips]{graphicx}

\begin{document}

\title{The solar \BaII\  4554 \AA\  line as a Doppler diagnostic:
       NLTE analysis in 3D hydrodynamical model}
\author{N. G. Shchukina\inst{1} \and V.L. Olshevsky\inst{1}  \and E. V.~Khomenko\inst{2}$^,$ \inst{1} }
\institute{Main Astronomical Observatory, NAS, 03680 Kyiv,
Zabolotnogo str. 27, Ukraine \and Instituto de Astrof\'{\i}sica de
Canarias, 38205, C/ V\'{\i}a L{\'a}ctea, s/n, Tenerife, Spain
\email{khomenko@iac.es} } \offprints{E.V.~Khomenko}
\date{Received date; accepted date}

\abstract
{}
{The aim of this paper is  to analyse the validity of the
Dopplergram and $\lambda$-meter techniques for the Doppler
diagnostics of solar photospheric velocities using the \BaII\ 4554
\AA\ line.}
{Both techniques are evaluated by means of NLTE radiative transfer
calculations of the \BaII\ 4554 \AA\ line in a three-dimensional
hydrodynamical  model of solar convection. We consider the cases
of spatially unsmeared profiles and the profiles smeared to the
resolution of ground-based observations.}
{We find that: (i) Speckle-reconstructed Dopplergram velocities
reproduce the ``true'' velocities well at heights around 300 km,
except for intergranular lanes with strong downflows where the
velocity can be overestimated.
(ii) The $\lambda$-meter velocities give a good representation of
the ``true'' velocities through the whole photosphere, both under the
original and reduced spatial resolutions. The velocities derived
from the inner wing of smeared \BaII\ 4554 \AA\ line profiles are
more reliable than those for the outer wing. Only under high spatial
resolution does the inner wing velocities calculated in intergranular
regions give an underestimate (or even a sign reversal) compared
with the model velocities.
(iii) NLTE effects should be taken into account in modelling the
\BaII\ 4554 \AA\ line profiles. Such effects are more pronounced
in intergranular regions. }
{Our analysis supports the opinion that the Dopplergram technique
applied to the \BaII\ 4554 \AA\ line is a valuable tool for the
Doppler diagnostics of the middle photosphere around 300 km. The
$\lambda$-meter technique applied to this line gives us a good
opportunity to ``trace'' the non-thermal motions  along the whole
photosphere up to the temperature minimum and lower chromosphere.}


\authorrunning{Shchukina et al.}
\titlerunning{\BaII\  4554 \AA\  line as  Doppler diagnostics}

\maketitle


\section{Introduction}

First studies of the barium lines in stellar spectra started in the
1930s. One of the examples of these earlier studies is by
\citet{bid:keen:1951}, who pointed out that the anomalous  strength of
the \BaII\  4554 \AA\ line in some stars could be explained by a
deviation from local thermodynamic equilibrium (NLTE).
%
At the present time stellar  studies are aimed at the abundance
determination of barium isotopes
using one-dimensional (1D) plane--parallel atmospheric  models and the
NLTE assumption
 \citep{gigas:1988, mash:gehr:2000, mash:etal:2003,
mash:etal:2008, mash:zhao:2006, sho:hau:2006}.
Such studies play important role in estimating the yields of s-
and r-processes in the nucleosynthesis of heavy elements in the
Galaxy.


Barium lines in the solar spectrum have been studied since 1960s.
 Until the mid-seventies most the studies aimed at
determining the solar barium abundance  using a 1D$+$LTE approach
\citep{gold:mul:all:1960, lamb:warn:1968, hol:mul:1974}. The only
exception was the publication of \citet{tand:1964,
tand:smyth:1970}, who had shown the importance of NLTE effects for
the formation of the Ba\,{\sc ii}  4554 \AA\, line. Later,
\citet{rutt:1977, rutt:1978, rutt:milk:1979} tackled in detail
the NLTE Ba\,{\sc ii}  line formation problem. Empirical analyses
of centre-to-limb observations by \citet{rutt:1978} have shown
that the  source function of the Ba\,{\sc ii}  4554 \AA\, line
deviates significantly from the Planck function.
The effects of partially  coherent scattering have also to be taken into
account in order to reproduce
observations away from the disc centre.
The hyperfine structure and isotopic shift play a very important
role in the analysis of this line as well.

For several reasons, a new debate has arisen in the recent
literature on the solar \BaII\ 4554 line.
Firstly, a more realistic representation of the solar atmosphere
by three-dimensional (3D) hydrodynamical simulations has become
available. Relaxing the constraints of the plane--parallel (1D)
modelling a new LTE abundance analysis by \citet{asp:etal:2005}
resulted in a photospheric barium abundance of $A_{\rm Ba}=2.17$,
close to the meteoritic value $A_{\rm Ba}=2.16$. While the recent NLTE
analysis by \citet{olsh:etal:2008} based on a 3D model lowers the value of
$A_{\rm Ba}$ strictly to the meteoritic one.
Note that the classical
1D-approach resulted in a rather wide spread of the solar barium
abundance  $A_{\rm Ba}=2.09$ \citep{ross:aller:1976},
$A_{\rm Ba}=2.13$ \citep{grev:1984}, and
$A_{\rm Ba}=2.40$ \citep{rutt:1978}.

Secondly, according to the atlas of the ``Second Solar spectrum''
\citep{gand:2002}, the linear polarization $Q/I$ of this line close to
the limb ($\mu=0.1$) is very strong ($0.6\%$). The recently
published theoretical investigation on the role of resonance
scattering and magnetic fields in the polarization signals of
both \BaII\  4934 \AA, and 4554 \AA\, resonance lines by
\citet{bell:etal:2007} has demonstrated their importance for the
measurements of weak magnetic fields on Sun and stars.

Finally, several properties of the \BaII\ 4554 \AA\ line have
drawn attention to it as a diagnostic tool for the velocity field
of the solar atmosphere. Due to the large atomic weight of barium
(137.4 a.u.), one might expect low sensitivity of the line opacity
to temperature variations and  line-width insensitivity to thermal
broadening. In addition, the \BaII\  4554 \AA\ line has steep
wings and a deep core.  In a standard 1D model the core of
this line is formed around 700 km in the chromosphere while the
wings are photospheric \citep{olsh:etal:2008, sutt:etal:2001}. As
a result, the \BaII\ 4554 \AA\ line gives an excellent opportunity
to ``trace''  non-thermal motions (granulation and
supergranulation velocity field and waves) throughout photosphere and  even in the lower chromosphere. \citet{noy:1967}
was one of the first to draw attention to the Doppler diagnostic
potential of this line. Later, \citet{rutt:1978} confirmed the
\BaII\ 4554 \AA\ line to be a perfect tool for investigating the
velocity structure of the solar photosphere and lower
chromosphere. Recently, \citet{sutt:etal:2001} have made the first
serious attempt to use the \BaII\ 4554 \AA\ line for mapping the
line-of-sight (LOS) velocities (Dopplershift map) of different
structures in the solar photosphere. They presented observations
with the Dutch Open Telescope (DOT) testing the Dopplergram capability
of narrow-band (80 m\AA) Lyot  filter \citep{sko:etal:1976}
imaging the solar surface in the wings of this  line in
combination with speckle reconstruction. The \BaII\ 4554 \AA\ line
is found to be an excellent tool for high-resolution Doppler
mapping.

Summarizing all the above, the \BaII\ 4554 \AA\ line provides a
valuable diagnostic tool for the solar and stellar atmospheres.
It is thus extremely important to investigate carefully the
validity of the different data-processing techniques and
interpretations applied to this line to obtain information on
physical conditions in the solar atmosphere.
Our paper presents an example of such an investigation. Below we
analyse two techniques used for recovering the solar velocity field
from observations in this line. The aim of our analysis is to
consider the advantages and disadvantages of these techniques and
to evaluate to what extent the LOS velocities provided by them
give a correct measure of the solar values.
We keep in mind that, besides the techniques themselves, there is
another important source of uncertainties  that can affect
Doppler diagnostics. The ground-based observations are typically
affected by the Earth's atmospheric turbulence (seeing) and
instrumental effects due to light diffraction on  the telescope
aperture (the finite spatial resolution of the telescope), the
finite instrumental width of the filters used, stray light, etc.
So we tackle the problem taking into account seeing and
instrumental effects.
The first technique analysed in this paper is a 5-point
Dopplergram method used by \citet{sutt:etal:2001}. The second
technique is known as a $\lambda$-meter, first proposed by
\citet{steb:good:1987}. In subsequent years it has been adopted by
several researchers  \citep[see][and more references therein]
{kostik:khomenko:2007, kostyk:shshu:2004,
khomenko:kostyk:shchu:2001} for spectral observations of different
Fraunhofer lines.

The organization of this paper is as follows.
Subsections~2.1 and 2.2 describe the 3D hydrodynamical model, atomic
data and numerical methods needed for  the NLTE radiative transfer
calculations with the barium atomic model. In Subsections~2.3 and 2.4
we discuss the procedure used to simulate spatial smearing of the
two-dimensional maps of the synthetic \BaII\ 4554 \AA\  line
profiles. We focus on seeing and instrumental effects. We
establish theoretical calibration dependences between the
granulation contrast and the Fried parameter (which specifies the
characteristic size of atmospheric turbulence cells) based on a
3D-approach. Subection~2.5 defines artificial datasets employed in
the paper while Section~3 presents obsevations. In Section 4 we
discuss the validity of the 5-point Dopplergram technique used to
obtain LOS velocities from the speckle-reconstructed observations
of the \BaII\ 4554 \AA\ line. Section~5 presents the results for
the $\lambda$-meter method. We discuss the granulation velocity field
that one would expect from observations of this line under perfect
spatial resolution (Subsection~5.1) and under different seeing
conditions of the ground-based observations (Subsection~5.2). An
extra point of particular interest has been to establish heights
from which the information on the velocity and intensity variations
originate. We discuss this problem in Subsection~5.3. Finally,
Section 6 presents our conclusions, while the Appendix gives a
brief description of our NLTE modelling with emphasis on the NLTE
mechanisms of formation of the  \BaII\ 4554 \AA\  line. We show
population departure coefficients, NLTE source functions and
profiles of this line to illustrate the difference in NLTE results
for granules and intergranules.

\section{Method}

\subsection{3D atmospheric model}
 \label{sec:3D_model}

We use a 3D snapshot from realistic radiation hydrodynamical
simulations of solar convection \citep{stein:nord:1998,
asp:etal:1999, asp:etal:2000, asp:etal:2000b}. This simulation is
based on a realistic equation of state, opacities and detailed
radiative transfer. The size of the simulation box is  $6
\times6 \times 3.8$~Mm, with 1.1 Mm being located above the
continuum optical depth equal to one. To reduce the amount of
time-consuming NLTE radiative transfer calculations, the snapshot
was interpolated from the original resolution of 200 grid points
in the horizontal direction to a coarser resolution of 50 grid points.
At the same time, we increased the resolution in the vertical
direction, taking only the upper 1.1 Mm part of the snapshot
between $-200$~km and 900~km and interpolating from 82 to 121 grid
points. Thus, the final 3D model has $50\times 50 \times 121$ grid
points or $50\times 50$ 1D models.

It was concluded from previous studies that the 3D model used here
performs very satisfactory in terms of spectral line formation
both for line shape and asymmetries \citep{asp:etal:1999,
asp:etal:2000, asp:etal:2000b, asp:etal:2004, Shchukina:etal:2005,
Shchukina:Trujillo:2001, Shchukina:Trujillo:2009,
trujillo:etal:2004}. The model reproduces all the main features of
solar convective velocities and intensities
\citep{kostyk:shshu:2004}. It reproduces as well the
root-mean-square (r.m.s.) contrast of solar granulation at the
6301 \AA\ continuum wavelength (as obtained by HINODE),
centre-to-limb variation of continuum intensity, and the polarization
of the solar continuum \citep{trujillo:shchu:2008}.

\begin{figure}
\includegraphics[width=9cm]{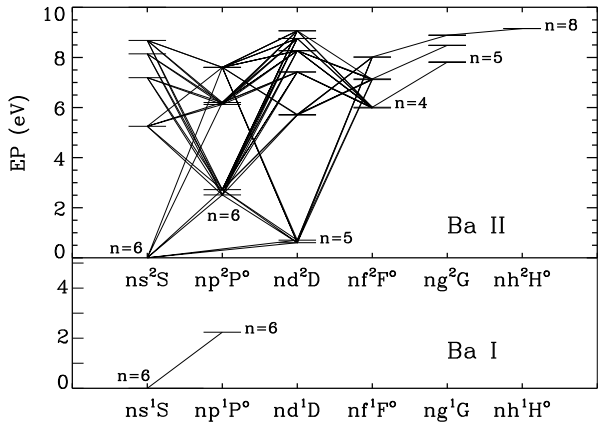}
\caption{Grotrian diagram of an atomic model for for Ba\,{\sc i}
+\BaII. Solid lines indicate radiative bound--bound transitions.
The \BaII\ $\lambda$4554 \AA\ line arises from the
$6s\, \rm{{^2}S}_{1/2} - 6p\, \rm{{^2}P}_{3/2}$ transition.} \label{fig:atom}
\end{figure}

\begin{figure}
\centering
\includegraphics[width=8.5cm]{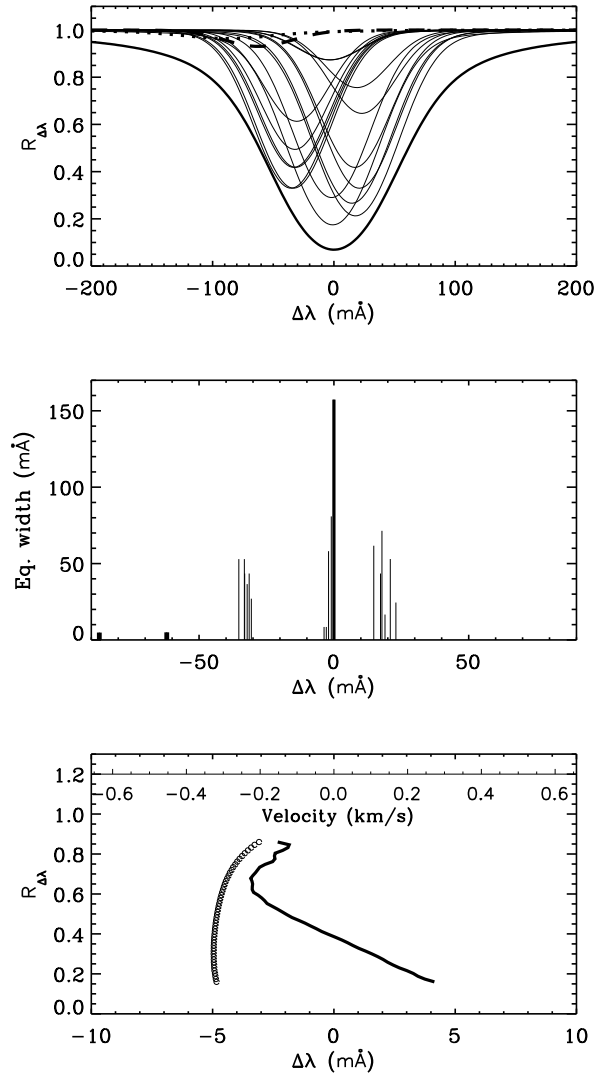}
\caption{Hyperfine and isotopic structure of the Ba \,{\sc ii}
4554 \AA\  line. {\it Top}: splitting profile patterns  (thin
solid lines) and resulting profile (thick solid line)  in the 1D
MACKKL model \citep{malt:etal:1986}. Two blends are shown by
dotted (Cr\,{\sc i} 4553.945 \AA) and dashed (Zr\,{\sc ii}
4553.970 \AA) lines. {\it Middle}: equivalent widths and isotopic
shift of the patterns. {\it Bottom}:  bisector due to isotopic
shift (small open circles) calculated in the MACKKL model and
bisector of the spatially averaged profile (3D model) caused by
the granular velocity field (solid line).} \label{fig:hyper}
\end{figure}
%

\subsection{Spectral synthesis}

According to \citet{rutt:1978, rutt:milk:1979, olsh:etal:2008}
NLTE is the crucial factor that should be taken into account when
calculating the \BaII\ 4554 \AA\ solar line profile. Here we
considered the NLTE barium line formation problem neglecting the
effects of horizontal radiative transfer (1.5D approximation). A
self-consistent solution of the kinetic and radiative transfer
equations has been obtained with an efficient multilevel transfer
code ``NATAJA''  developed by \citet{Shchukina:Trujillo:2001} to
facilitate NLTE radiative transfer simulations with very complex
atomic models. Before the code was successfully  used for NLTE
interpretation of iron, oxygen, titanium, and strontium solar
spectra \citep{Shchukina:Trujillo:2001, Shchukina:Trujillo:2009,
Shchukina:etal:2005, trujillo:shchu:2007, trujillo:etal:2004,
kostyk:etal:2006, kostyk:shshu:2004, khomenko:kostyk:shchu:2001}.
The code is based on iterative methods for radiative transfer
calculations \citep[see][and more references
therein]{trujillo:fabiani:1995, Socas:Trujillo:1997} that allow a
fast and accurate solution of NLTE transfer problems.

Our atomic model includes 40 energy levels of Ba\,{\sc i} and
\BaII\ (see Fig.~\ref{fig:atom}). Note that all levels with
hyperfine structure (HFS) were treated as a single level, \ie\ for
the HFS sublevels the  population departure coefficients $\beta$
were taken to be equal. The levels are interconnected by 99 bound--bound
and 39 bound--free radiative transitions. All levels are coupled
via collisions with electrons. The atomic model and atomic data,
including oscillator strengths, bound--free cross-sections,
electron collisional  rates, etc., are described in detail by
\citet{olsh:etal:2008}.

The departure coefficients $\beta$ found from the
self-consistent solution of the kinetic and radiative transfer
equations were used as input to carry out  the formal solution of
the radiative transfer equation for the \BaII\ $\lambda$4554 \AA\
line.
At this step of the solution  we took
into account the hyperfine structure (HFS) and isotopic
shift of the \BaII\ $\lambda$4554 \AA\ line (see
Fig.~\ref{fig:hyper}). Note that because of non-zero nuclear spin
only  two isotopes ($\rm{{^{135}}Ba}$ and $\rm{{^{137}}Ba}$) have
HFS-splitting. The energies of HFS-sublevels are calculated according
to a formula given by Radzig and Smirnov (1985). The hyperfine
structure constant $A$ needed for these calculations was taken
from \citet{rutt:1978}. We neglected the interaction of electrons
with a nuclear electric quadrupole momentum of because this
effect is rather small. The isotopic shift was derived using mass
shift and field shift constants  from \citet{ber:etal:2003} and
mean square nuclear radii from \citet{sak:tan:2001}. We
synthesized the \BaII\ $\lambda$4554 \AA\ line profiles employing
the isotopic abundance ratio \citep{radzig:smirnov:1985} for 17
sub-components. We included  2 blends (Cr\,{\sc i} 4553.945 \AA\
and Zr\,{\sc ii} 4553.970 \AA) and 6 spectral lines of other
elements observed in the far wings of this line. The upper panel of
Fig.~\ref{fig:hyper} shows that these blends produce only minor
effects on the synthesized profile.

We calculated the emergent intensities $I_{\Delta \lambda}$ along
the \BaII\ 4554 \AA\ line profile for the set of ${\Delta
\lambda}$ wavelength points and for every $(x_i,y_i)$ vertical
column of the 3D snapshot corresponding to the solar disc centre
$\mu = cos \Theta =1$ ($\Theta$ being the heliocentric angle). The
profiles were normalized to the mean continuum intensity
$\left<I_c\right>$ averaged over the snapshot. The continuum
intensity was also used as a criterion to separate granular and
intergranular regions. The $(x_i,y_i)$-grid points with continuum
intensity greater than $\left<I_c\right>$ were taken as granules
(and the opposite for intergranules).

Since we are doing calculations for the disc centre, we assume
complete frequency redistribution (CRD) for the \BaII\ 4554 \AA\
line. According to \citet{rutt:1978, rutt:milk:1979} the effects
of partial frequency redistribution (PRD) increase towards the limb,
where the frequency-dependent wing source structure becomes
noticeable.  At the disc centre the differences between the PRD
and CRD profiles of this line are small  (1\% of the continuous
intensity in the core) so we can safely neglect the effects of
PRD.

We do not use any ad hoc parameters such as micro- or
macroturbulence for the synthesis since the profiles are broadened
in a natural way by the velocity field existing in the 3D model.
The damping constant for the barium lines was determined as
the sum $\gamma={\gamma}_6+{\gamma}_{\rm rad}$ of van der Waals
collisional broadening ${\gamma}_6$ by neutral hydrogen and helium
atoms and radiative broadening ${\gamma}_{\rm rad}$. The other
collisional broadening processes (Stark broadening, quadrupole
broadening) are negligible \citep{rutt:1978}. We employ the
${\gamma}_6$ based on a theory in which the van der Waals
potential is replaced by a Smirnov--Roueff potential for the close
interactions \citep{der:rens:1976}.
In view of the uncertainty of collisional damping, we treat  the
${\gamma}_6$-value  as a free parameter by introducing the usual
enhancement factor $E$.
 In this study we use
$E=1.3$ and $A_{\rm Ba}=2.16$ derived earlier by
\citet{olsh:etal:2008} from the \BaII\ 4554 \AA\, 1.5D NLTE line
modelling in the same 3D snapshot. For these values of $E$ and
$A_{\rm Ba}$ the authors obtained excellent agreement between
the spatially averaged synthetic profile and  the observed one taken
from Li\'ege atlas \citep{liege}.
We understand the limitations of the Smirnov--Roueff potential
approximation for the close interactions. These limitations might
be overcome using the semi-classical theory of
\citet{barklem:omara:1998}. The introduction of the broadening for
P-D states given by these authors can be important for a  proper
modelling of the \BaII\ 4554 \AA\ line, particularly, for any
accurate calculation of the polarization Stokes amplitudes $(Q/I)$
produced by scattering processes in the solar atmosphere. The
elastic collisions with neutral hydrogen atoms are known to be
efficient in modifying the atomic level population of long-lived
levels, like \BaII\ $\rm{5{^2}D}$. Recent investigation of the
role of collisional depolarization of the \BaII\ 4554 \AA\ line in
the low chromosphere by \citet{derouich2008} shows that this line
is clearly affected by isotropic collisions with neutral hydrogen
atoms through the effect of collisions on the $\rm{5{^2}D}$ level.
However, the impact  of depolarizing elastic collisions on the
emergent intensity $(I)$ profiles in a weekly anisotropic medium
like the solar photosphere is expected to be negligible
\citep[e.g., see the book by][]{landi:landolfi:2004}.

%

In Appendix \ref{section:NLTE} we
summarize the NLTE effects on the formation of the  \BaII\ 4554 \AA\
line in the 3D snapshot and the compare NLTE and LTE calculations.

Below, we use the concept of the Eddington--Barbier height
of line formation, \ie\ we evaluate the height $H$ where the line
optical depth at a given wavelength point ${\Delta \lambda}$ is
unity: $\tau({\Delta \lambda}) =1$. We then assume that the
information about vertical velocity $V_z$ at wavelength point
${\Delta \lambda}$ originates from height $H$. The reader should
be aware of the practical limitations of such a concept
\citep[see][]{sanchez:etal:1996}. However, for the purposes of
statistical quantification of the errors in Doppler diagnostic
methods it is reasonable  and convenient to use the concept of the
Eddington--Barbier height of formation.  We assumed that
two-dimensional maps of the synthesized intensity profiles
represent observations in the case of perfect seeing conditions
and no instrumental effects. We then use such ``perfect
observations'' to compute Dopplergrams and $\lambda$-meter
velocities and to compare them to the ``true'' snapshot velocities
at the corresponding heights $H$.

\subsection{Spatial smearing}
\label{sec:spatial_smearing}

 Since ``perfect observations'' are only of
theoretical interest to an observer, we study here the case of
several degrading effects applied to the profiles. Observed images
are degraded because of the Earth's atmospheric turbulence
(seeing) and light diffraction b ythe telescope aperture (the finite
spatial resolution of the telescope). There are other factors,
such as stray light, but we expect that the observations are
corrected for them during the reduction process.

Mathematically, the Fourier transform of the image registered by
detector is related to the Fourier transform of the original image
via the modulation transfer function (MTF) as follows:
\begin{equation}
  I(\vec{f}) = O(\vec{f}) \left<\tau(\vec{f})\right>,
\end{equation}
where $\vec{f}$ is the  spatial frequency in ($\rm {rd^{-1}}$), and
$\left<\,\right>$ denotes the ensemble average. $O(\vec{f})$ is the
Fourier transform of the observed object,  $I(\vec{f})$ is the
Fourier transform of the image registered with a telescope camera,
$\tau(\vec{f})$ is the MTF of the atmosphere and telescope.
Following \citet{fried:1966}, for exposures longer than the
characteristic lifetime of the atmospheric turbulence, the MTF is
given by:
\begin{equation}
\left<\tau(f)\right>_{LE} = \tau_0(f) \exp(-3.44(\lambda
f/R_0)^{5/3}),
\end{equation}
where $f=|\vec{f}|$, and $\tau_0(f)$ is the autocorrelation
function of the telescope pupil:
\begin{equation}
\tau_0(f) = {2\over \pi} \left[{\rm cos}^{-1}\left({\lambda f
\over D}\right) - {\lambda f \over D}\left(1 - \left({\lambda f
\over D}\right)^2\right)^{1/2} \right],
\end{equation}
if ${\lambda f / D} \le 1$. In such an interpretation only two
parameters define the resulting image quality: telescope diameter
$D$ and Fried's parameter $R_0$ \citep{fried:1966, korff:1973}.
The latter parameter depends on seeing conditions and describes
characteristic size of atmospheric turbulence cells.

At each wavelength, the original 2D intensity maps were
Fourier-transformed and multiplied by the MTF calculated for a known
telescope diameter and Fried's parameter $R_0$. An inverse Fourier
transform gives us the images registered by detector, \ie\ the
``observed'' images, affected by the diffraction by the telescope
aperture and the seeing effects.

\subsection{Determination of Fried's parameter}
 \label{sec:fried}

In order to make a direct comparison between the synthetic and
observed spectra we have to know Fried's parameter $R_0$. There is
no direct way to measure such a parameter in observations.
\citet{ric:etal:1981} proposed determining $R_0$ from the
observed r.m.s. continuum contrast $\delta
I_{\rm r.m.s.}$ of solar granulation. Their computations are based
on using an analytical form for the power spectrum of the intensity
distribution at $\lambda=5050$~\AA\ obtained by
\citet{ric:aime:1979} from speckle interferometric observations of
the solar granulation. To establish calibration dependence between
$R_0$  and $\delta I_{\rm r.m.s.}$ at other wavelengths
\citet{ric:etal:1981} used simplified assumptions that the solar
photosphere behaves as a  blackbody, and that the differences in
formation heights of continuum radiation as a function of
wavelength can be ignored.

\begin{figure}
\centering
\includegraphics[width=9cm]{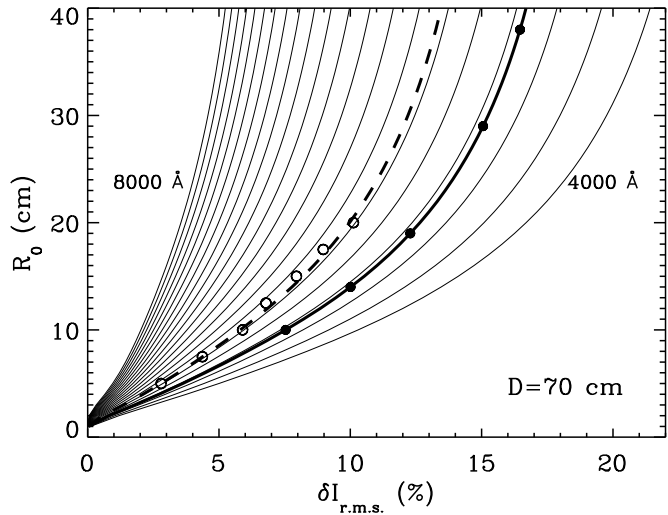}
\caption{Root-mean-square continuum contrast $\delta I_{\rm
r.m.s.}$ as a function of Fried's parameter $R_0$ for different
wavelengths and for telescope diameter $D=70$~cm. $\delta I_{\rm
r.m.s.}$ curves were calculated using the 3D radiative transfer formal
solution for continuum intensity obtained by
\citet{trujillo:shchu:2008} in the 3D HD model. Dashed and thick
solid line: $\delta I_{\rm r.m.s.}$ at $\lambda=5050$~\AA\ and
$\lambda=4554$~\AA, respectively. Thin solid lines: $\delta I_{\rm
r.m.s.}$ for wavelengths from $\lambda=4000$~\AA\ till
$\lambda=8000$~\AA\ plotted with step 200 \AA. Filled circles
correspond to $R_0=10, 14, 19, 29, 38$~cm. Open circles represent the
data for $\lambda=5050$~\AA\ obtained by \citet{ric:etal:1981}
using an empirical approach.} \label{fig:contrasts_3D}
\end{figure}

Here we propose another way to set up calibration curves for
determination of $R_0$ from granular contrast in
observations. Our calculations use a 3D radiative
transfer solution for the solar continuum intensity obtained by
\citet{trujillo:shchu:2008} in the same 3D snapshot as was
employed above for the \BaII\ profile synthesis (see
Sec.~\ref{sec:3D_model}). Note that such a 3D approach gives a
possibility of avoiding the simplification used by \citet{ric:etal:1981}.
The theoretical calibration curves for the wavelengths between
4000~\AA\ and 8000~\AA\ are plotted in
Figure~\ref{fig:contrasts_3D}.

In order to verify the 3D approach we compare the empirical
calibration curve at $\lambda=5050$~\AA\ taken from figure 1 of
\citet{ric:etal:1981} with our theoretical curve. The analysis of
the results presented in Fig.~\ref{fig:contrasts_3D} allows us to
conclude that at $\lambda=5050$~\AA\ both the empirical (open
circles) and the theoretical (dash line) curves in fact coincide.
Keeping in mind such a good agreement we use our theoretical
3D approach for determination of  $R_0$ in
Section~\ref{sec:obs}.

%
%

\begin{figure}
\centering
\includegraphics[width=8.5cm]{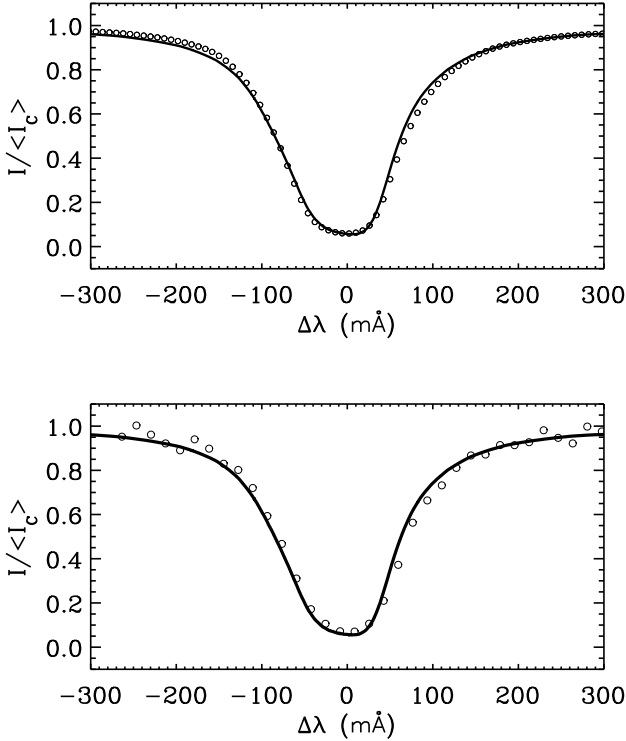}
\caption{Calculated and observed intensity disc-centre profiles of
the \BaII\ 4554 \AA\ line. Solid line: the spatially averaged NLTE
profile for the 3D model. Small open circles: the profile from the
Li\`ege Atlas ({\it top}); spatially averaged profile obtained
from observations at the VTT with TESOS in 2006 ({\it bottom}).}
\label{fig:obs}
\end{figure}
%

\begin{figure}
\centering
\includegraphics[width=8cm]{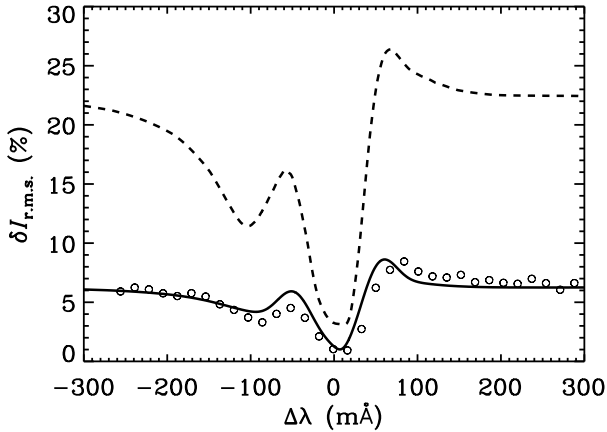}
\caption{Root-mean-square contrast at different wavelengths along
the  \BaII\ 4554 line profile. Dashed line: $\delta I_{\rm
r.m.s.}(\delta\lambda)$ in unsmeared synthetic data. Thick solid
line: $\delta I_{\rm r.m.s.}(\delta\lambda)$ in the synthetic data
smeared using Fried's parameter $R_0=9$~cm. Circles correspond to
the observations at the VTT with TESOS in 2006.}
\label{fig:contrasts}
\end{figure}
%

\subsection{Definition of artificial datasets}
\label{subsection:nomenclature}

Here we define the nomenclature for the artificial datasets to be
used in the rest of the paper.

{\sc DOT-like data.} \citet{sutt:etal:2001} used an Irkutsk barium
filter \citep{sko:etal:1976} installed on the Dutch Open Telescope to
obtain Dopplergrams in the \BaII\ 4554 \AA\ line. We have calculated
Dopplergrams from the synthetic profiles in order to model the
results of \citet{sutt:etal:2001}. To do that, we smeared the
original profiles using the DOT's diameter of $D=48$ cm. In
addition, spectral smearing was performed by convolving the
profiles in wavelength with a Gaussian of FWHM (full width half
maximum) of 80 m\AA, corresponding to the barium filter spectral
profile.  The Dopplergrams were constructed as a second-order
polynomial fit to the five points on the line profile situated at
$\Delta\lambda=-70$, $-35$, 0, 35 and 70 m\AA. The observations of
\citet{sutt:etal:2001} were speckle-reconstructed, so we assumed
that the influence of the Earth's atmosphere was minimized, and that no
smearing with the Fried parameter was applied.

{\sc VTT-like data.} The second type of artificial data represents
observations done with the German Vacuum Tower Telescope (VTT) and
Triple Etalon SOlar Spectrometer \citep[TESOS;][]{trit:etal:2002}.
The synthetic profiles were convolved using the telescope diameter
$D=70$ cm, and a spectral bandwidth of TESOS of
FWHM=15 m\AA. The Fried parameters in this case ranged
from $R_0=10$~cm (medium seeing) to $R_0= 38$~cm (excellent
seeing).
According to Fig.~\ref{fig:contrasts_3D}
(thick solid line and filled circles)
the corresponding root-mean-square continuum contrast $\delta
I_{\rm r.m.s.}$ of solar granulation
at  the  wavelength of the  \BaII\ 4554 \AA\ line
varies from 7.5\% to
16.5\%.
%


\section{Comparison to observations}
\label{sec:obs}

In this paper we used observations of the \BaII\ 4554 \AA\, line,
obtained in September 2006 by E. Khomenko, M. Collados and R.
Centeno at the 70-cm German Vacuum Tower Telescope (VTT) at the
Observatorio del Teide in Tenerife with the help of Triple Etalon
SOlar Spectrometer \citep[TESOS;][]{trit:etal:2002}. The quiet Sun
region close to the disc centre was observed, the presence of the
magnetic activity was controlled by the simultaneous observation
with the Tenerife Infrared Polarimeter II \citep[TIP
II;][]{coll:2007} in \FeI\ lines at 1.56 $\mu$m. We took a single
TESOS wavelength scan in \BaII\ 4554 \AA\, line made on September
2, which represents a series of 36 narrow-band filter images
(FWHM=15~m\AA) obtained with 10 m\AA\ interval along the 4554~\AA\
line profile. Each image was taken with a 500  $\mu$s exposure,
taking around 26 s to scan the complete line profile. The pixel
resolution of the TESOS camera was  0.089~arcsec. Due to the
instrument specifics, the TESOS field of view is circular. Taking
into account that the image quality towards its edge is worse we
have chosen a $256\times 256$ pixel square cut from the central
part of each image. By spatially averaging each of the 36 filter
images we got the mean intensity profile of the \BaII\ 4554 \AA\
line normalized to the mean continuum intensity of the Li\`ege
Atlas \citep{liege}.

Figure~\ref{fig:obs} shows observed and computed spatially
averaged profiles of the \BaII\ 4554 \AA\ line. The top panel of
Fig.~\ref{fig:obs} demonstrates that the profile computed using
the 3D snapshot is in a good agreement with the observed one taken
from the Li\'ege atlas \citep{liege}. The central part  of the
line  is reproduced very well while the fit for the outer red wing of the
computed profile is a little worse. The averaged line
profile obtained in 2006 with the TESOS instrument at the VTT is
shown in the bottom panel of Figure~\ref{fig:obs}. Again, both
computed and observed profiles agree well. The saw-tooth shape
of far wings of the observed profile arises because of  tuning
effects of the TESOS etalons \citep{trit:etal:2002}. We would like
to stress that such a good match of the observations is not based
on free parameter fitting but has been achieved self-consistently
through the NLTE plane--parallel modeling combining a realistic
atomic model and the 3D hydrodynamical snapshot.

Figure~\ref{fig:contrasts} shows the run of the r.m.s. contrast
with wavelength along the \BaII\ 4554 \AA\ line for three sets of
single-wavelength images: observed, smeared, and non-smeared
synthetic data. The synthetic profiles were smeared employing the
method developed in Section~\ref{sec:spatial_smearing}. We applied
the theoretical calibration curve at $\lambda=4554$~\AA\ shown in
Fig.~\ref{fig:contrasts_3D} (thick solid line) to define the Fried
parameter  from the VTT observations of the \BaII\ 4554 \AA\
line. We found that  during observations the r.m.s. continuum
contrast $\delta I_{\rm r.m.s.}$ at the wavelength of this line
was around 7\%. Thus, we estimate  $R_0$ to range
range between 9 and 10~cm. Synthetic data in
Fig.~\ref{fig:contrasts} are smeared using $R_0=9$~cm.

Seeing effects influence the r.m.s. contrast in the line wings,
while in the line core its influence is not so significant,
although the wing contrast is much higher than the contrast in the
line core. Two local maxima near $\Delta \lambda = \pm 50$ m\AA\
are not so high and sharp in the observed data as in the smeared
computed profiles. While at other wavelengths the agreement is
fairly good. Such blurring of the maxima may be due to the wider
bandwidth of TESOS filter system than given by
\citet{trit:etal:2002}.

Comparison of the synthetic and observed line profiles performed
in this section suggests that 3D atmospheric model of
\citet{asp:etal:2000} together with our NLTE calculations and
treating of atmospheric and instrumental influence, reproduce well the
average characteristics of the observed granulation pattern in Ba
II 4554 \AA\, line.

\section{Dopplergram technique}

\begin{figure*}
\centering
\includegraphics[width=14cm]{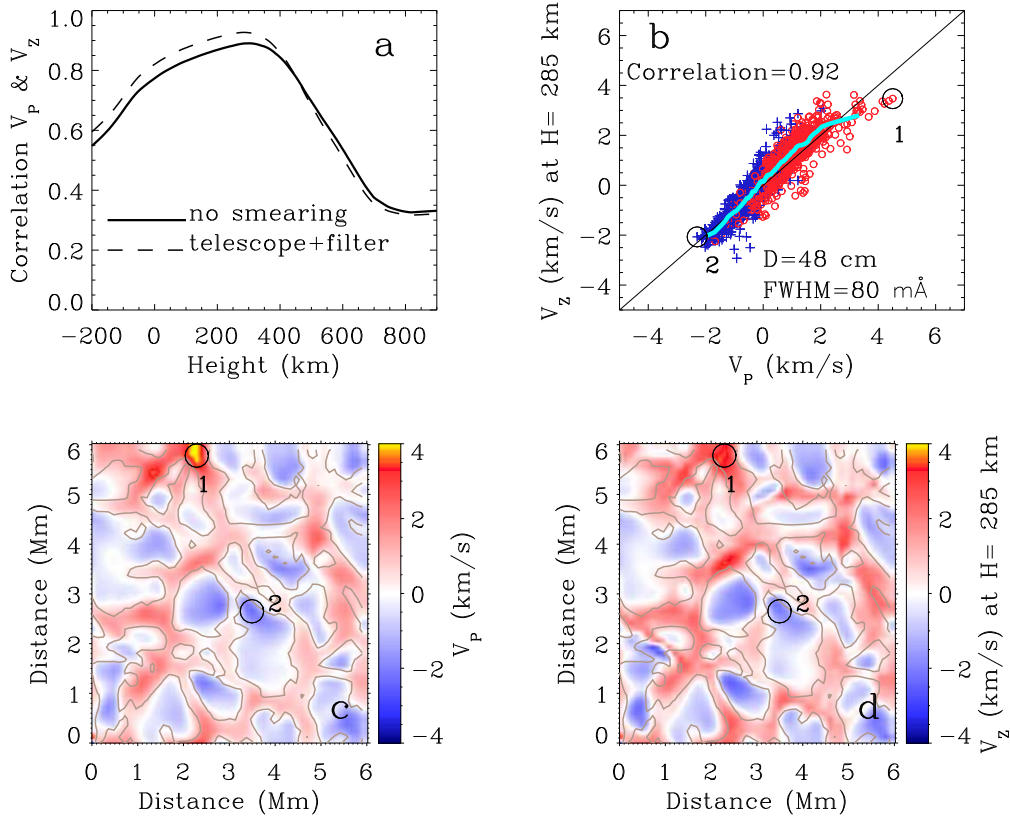}
\caption{Comparison of the snapshot vertical velocities $V_z$ with
the Dopplergram velocities $V_P$ inferred from the DOT-like data.
({\it a}): Correlation coefficients between $V_P$ and $V_z$ as a
function of height for the case of no smearing (solid line) and
telescope+filter smearing (dashed line). ({\it b}): Dopplergram
velocities $V_P$  vs. snapshot velocities  $V_z$. The latter were
taken at heights where the correlation between $V_P$ and $V_z$
reaches maximum. Blue crosses are granular points and red circles
are intergranular points. The light blue curve is the average over
bins with 100 surface points. ({\it c}): Map of the Dopplergram
velocities $V_P$. Circles with numbers (1), (2) correspond to the
spatial points where the emergent profiles have high red and blue
shifts, respectively. Those profiles are shown in
Figure~\ref{fig:parab_prof}. Red denotes downflows and blue,
 upflows. ({\it d}): Map of the snapshot velocities
$V_z$. } \label{fig:parab_map}
\end{figure*}

We discuss in this Section the range of validity of the 5-point
Dopplergram technique used by \citet{sutt:etal:2001} at the DOT
to obtain LOS velocities.
The comparison between the Dopplergram velocities from the
DOT-like data $V_P$ and the ``true'' snapshot velocities $V_z$ is
given in Figure~\ref{fig:parab_map}. The velocities $V_P$ were
corrected for the asymmetry caused by the hyperfine structure (the
bottom panel of Figure~\ref{fig:hyper}).

When applying the Dopplergram technique, it is important to
identify the height in the atmosphere from where the velocities
originate. We assumed that the information about velocities
at different points around the snapshot comes from nearly the same
heights. Then we calculated the correlation coefficient between
the maps of $V_P$ and $V_z$, taking the latter at different
heights, as shown in Figure~\ref{fig:parab_map}a. This correlation
tells us at what height the Dopplergram velocities
of the DOT-like data measure the real solar velocities. As we can
see, the correlation coefficient strongly depends on the
atmospheric height, for both the smeared and non-smeared data. The
maximum correlation reaches the value $\sim$0.9 at heights around
300 km. Thus, we can identify a rather narrow layer where the
Dopplergram velocities originate.

Figure~\ref{fig:parab_map}b gives the scatter plot of $V_P$ and
$V_z$, the latter taken at the height of maximum correlation (285
km). It demonstrates that in general the speckle-reconstructed
Dopplergram velocities correspond to the true snapshot velocities
rather well, with the standard deviation being less than 0.4 km/s.
The surface maps of $V_P$ and $V_z$ also agree rather well (panels
c and d). Thus, the Dopplergram velocities reproduce in many
details the ``true'' granulation velocity structure  existing at
heights around 300 km. Nevertheless, it can be seen from the maps
that at some locations in intergranular lanes the Dopplergram
velocities overestimate the snapshot velocities. These are the
locations with strong downflows. The difference between $V_P$ and
$V_z$ can be as high as 1 km/sec at these locations. Such an
excess manifests itself in Dopplergram velocity maps as patches of
enhanced brightness (Fig.~\ref{fig:parab_map}c, location marked by
(1)), being absent in the snapshot velocity maps.

\begin{figure}
\centering
\includegraphics[width=7cm]{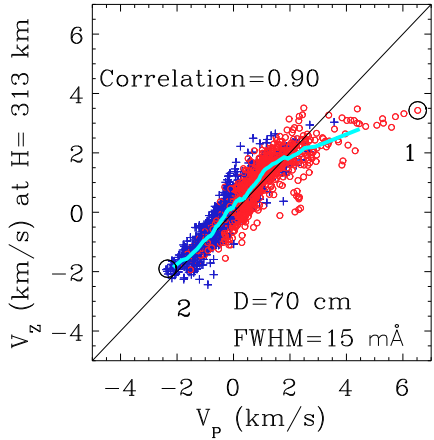}
\caption{Comparison of the snapshot vertical velocities $V_z$ with
the Dopplergram velocities $V_P$ inferred from the
``speckle-reconstructed'' VTT-like data.} \label{fig:parab_VTT}
\end{figure}

Interestingly, with the narrower  bandwidth and larger
diameter of the telescope than in the case of the DOT observations
the overestimation  of the Doppler velocities $V_P$ in
intergranular  areas with  strong downflows found from the
``speckle-reconstructed observations'' turns out to be appreciably
larger. We display such a case for VTT-like data  in
Figure~\ref{fig:parab_VTT}. As follows from the Figure, the
deviation reaches approximately 3 km/s.

\begin{figure*}
\centering
\includegraphics[width=6cm]{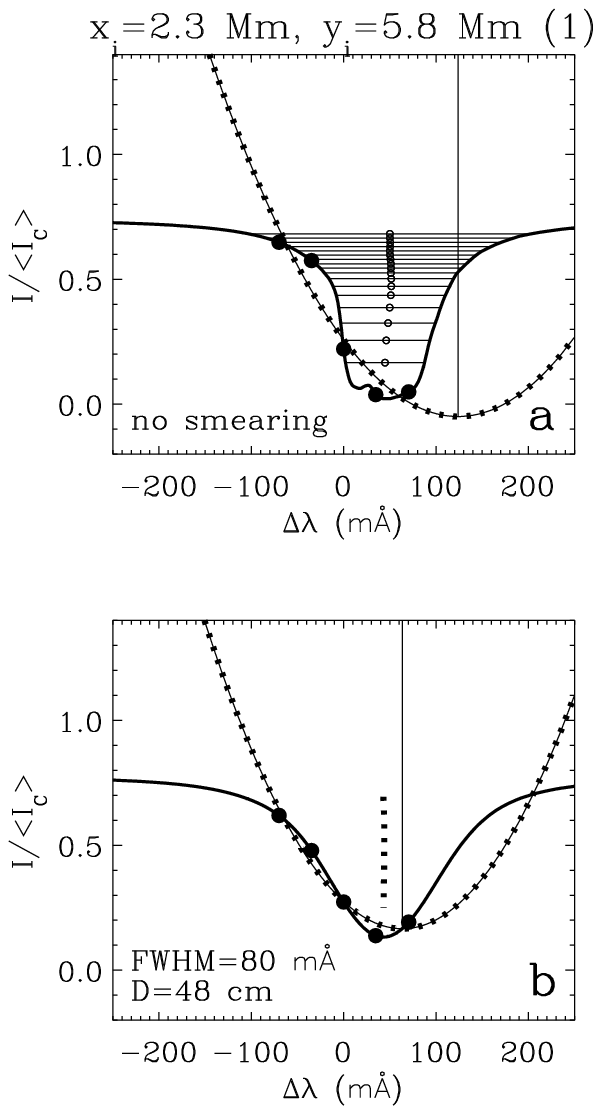}
\includegraphics[width=6cm]{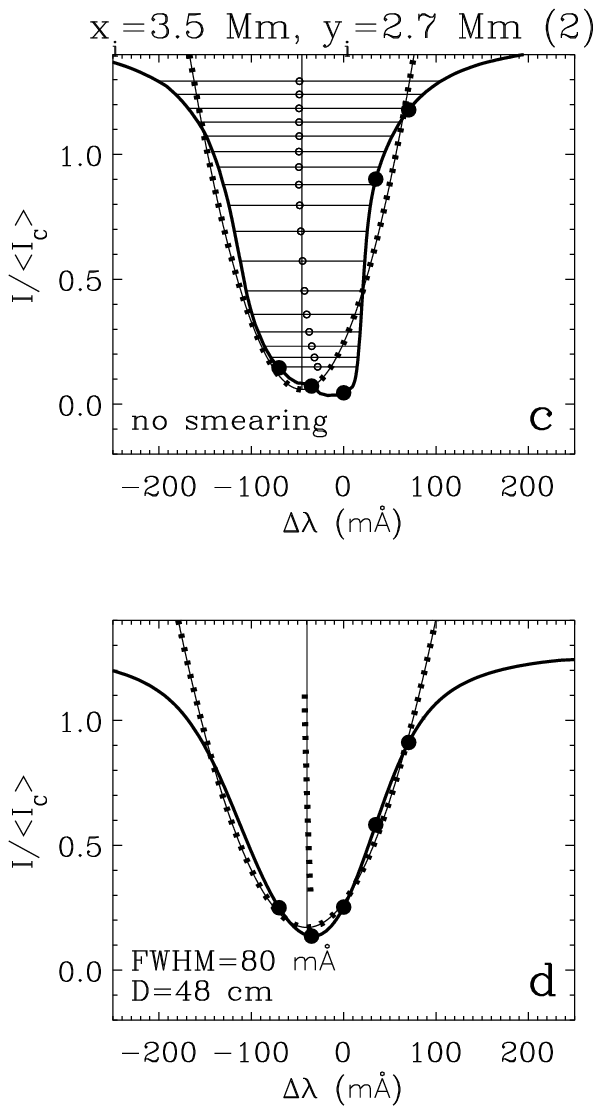}
\caption{ \BaII\ 4554 \AA\ line profiles at two selected locations
corresponding to a high downflow (left) and a high upflow (right),
marked by (1) and (2) in Figure\ref{fig:parab_map}. ({\it a, c}):
profiles without smearing. Thin lines with dots: parabolic fit to
the five wavelength positions $\Delta \lambda=-70$, $-35$, 0, 35
and 70 m\AA\ indicated  by black circles. Open circles: bisector
of the \BaII\  profile; thin vertical line: bisector of the
parabolic fit. The horizontal lines are full spectral line widths
$\Delta {\lambda}_W=\Delta {\lambda}_r-\Delta {\lambda}_b$
specifying the wavelength distances between equal intensity points
in the red and blue line wings. ({\it b, d}): same but for the
DOT-like smeared profiles. $x_i$ and $y_i$ are the locations of
the selected grid-points  in the 3D-snapshot.}
\label{fig:parab_prof}
\end{figure*}


Figure~\ref{fig:parab_prof} explains  why such bright  patches can
arise in speckle-reconstructed observations and why it happens
specifically in intergranular areas. It displays examples of the
parabolic fit to the five fixed wavelength points of the original
profiles (top) and DOT-like smeared profiles (bottom). The
profiles are taken at the spatial locations marked  (1) and (2) in
Figure~\ref{fig:parab_map}b, c, d. The profile marked (1)
originates from the area with a high intergranular downflow (left)
and the profile marked  (2) originates from the area with a high
granular upflow (right). According to Fig.~\ref{fig:parab_prof}a,
it seems impossible to measure correctly the velocity from the
weak twisted intergranular profile provided that the strong
redshift forces four of the five wavelength points used for
fitting to be situated in the blue wing. As a consequence, the
parabola turns out to be much more redshifted than the original
profile. So the Dopplergram velocity derived from the wavelength
position of the parabola bisector (thin vertical line) will be
considerably greater compared to the velocity inferred from the
bisector of the original profile (small open circles). Smearing
caused by the telescope and filter to some degree smoothes the
irregular shape of the intergranular profile and decreases the
redshift of the parabola (Fig.~\ref{fig:parab_prof}b).
Nevertheless, the velocity remains overestimated. Thus, we
conclude that the Dopplergram velocities calculated from the
speckle-reconstructed DOT-like data in intergranular lanes can be
appreciably greater than the real ones.

Contrary to the intergranular profile the deep and more symmetric
granular profile allows us to estimate the velocity from the
parabolic fit rather well (Fig.~\ref{fig:parab_prof}c). The
effects of smearing are less pronounced in this case
(Fig.~\ref{fig:parab_prof}d). So in granular areas with strong
upflows one can expect a reasonable agreement between the ``true''
velocities $V_z$ and velocities $V_P$ recovered from the DOT-like
data.

Summarizing the conclusion of this Section, the five-point
Dopplergram technique applied to the \BaII\ 4554 \AA\ line
profiles in  speckle-reconstructed DOT-like data is a valuable
tool for the diagnostic of the solar velocity field at heights
around 300 km. Only in intergranular lanes with strong downflows
can the velocity be overestimated producing artificially bright
points at the Dopplergram velocity maps like those obtained by
\citet{sutt:etal:2001}. These authors interpreted the bright
points as locations of the magnetic flux tubes. However, our
calculations show that at least part of such bright points may
be simply an artefact caused by the parabolic fit.

\section{$\lambda$-meter technique}

The $\lambda$-meter method of \citet{steb:good:1987} can be
considered as a form of the ``bisector shift'' technique proposed
earlier  by \citet{Kulander:Jefferies:1966} to evaluate the
atmospheric velocity field from the asymmetries in observed line
profiles. The $\lambda$-meter method deals with two line profile
parameters such as the line bisector and the full spectral line
width $\Delta {\lambda}_W$.

The displacement of a midpoint of a section of the line profile of
a certain width $\Delta {\lambda}_W$ is assumed to be a result of
the Doppler shift of the line opacity coefficient caused by the
non-thermal velocities in the layer where this section is formed.
According to \citet{steb:good:1987} the velocities and intensities
from progressively deeper sections of the line profile correspond
to progressively higher layers in the atmosphere. Following the
$\lambda$-meter procedure we introduced a set of line profile
widths  $\Delta {\lambda}_W$ ranging from 306 to 78 m\AA\ (see
Fig.~\ref{fig:parab_prof}). Due to the irregular shape of the
\BaII\ 4554 \AA\ line core, a significant number of synthetic
profiles never have a spectral line width lower than 78 m\AA. At
the same time shallow far wings of many  profiles do not allow us
to detect and analyse  velocities at widths above $\Delta
{\lambda}_W =306$~m\AA. The selected  $\Delta {\lambda}_W$ range
corresponds to the intensities of spatially averaged profile
varying between 0.86 and 0.16. The corresponding range of the mean
formation heights lies between $\sim -50$ and $\sim 500$~km.

\begin{figure*}
\centering
\includegraphics[width=14cm]{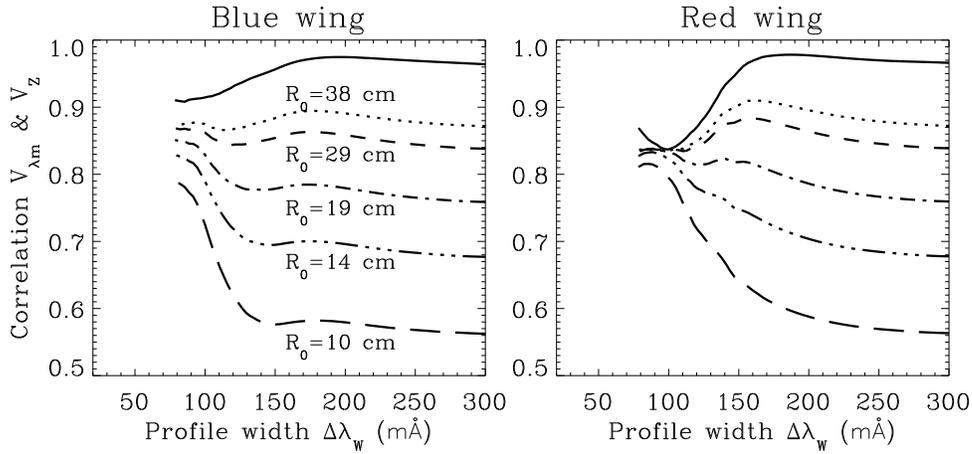}
\caption{Correlation coefficients between $\lambda$-meter
velocities $V_{\lambda M}$ and snapshot velocities $V_z$ as a
function of the line profile width $\Delta {\lambda}_W$. The
snapshot velocities are taken at heights $H$ where optical depth
at the corresponding wavelength $\tau(\Delta {\lambda}_{b(r)})$
equals unity (see Sect. 2.2). Each line corresponds to a different
value of $R_0$. The uppermost solid line refers to the case of no
smearing. Left panel: blue wing velocities; right panel: red wing
velocities. The VTT-like data are used.} \label{fig:lambda1}
\end{figure*}

\begin{figure*}
\centering
\includegraphics[width=14cm]{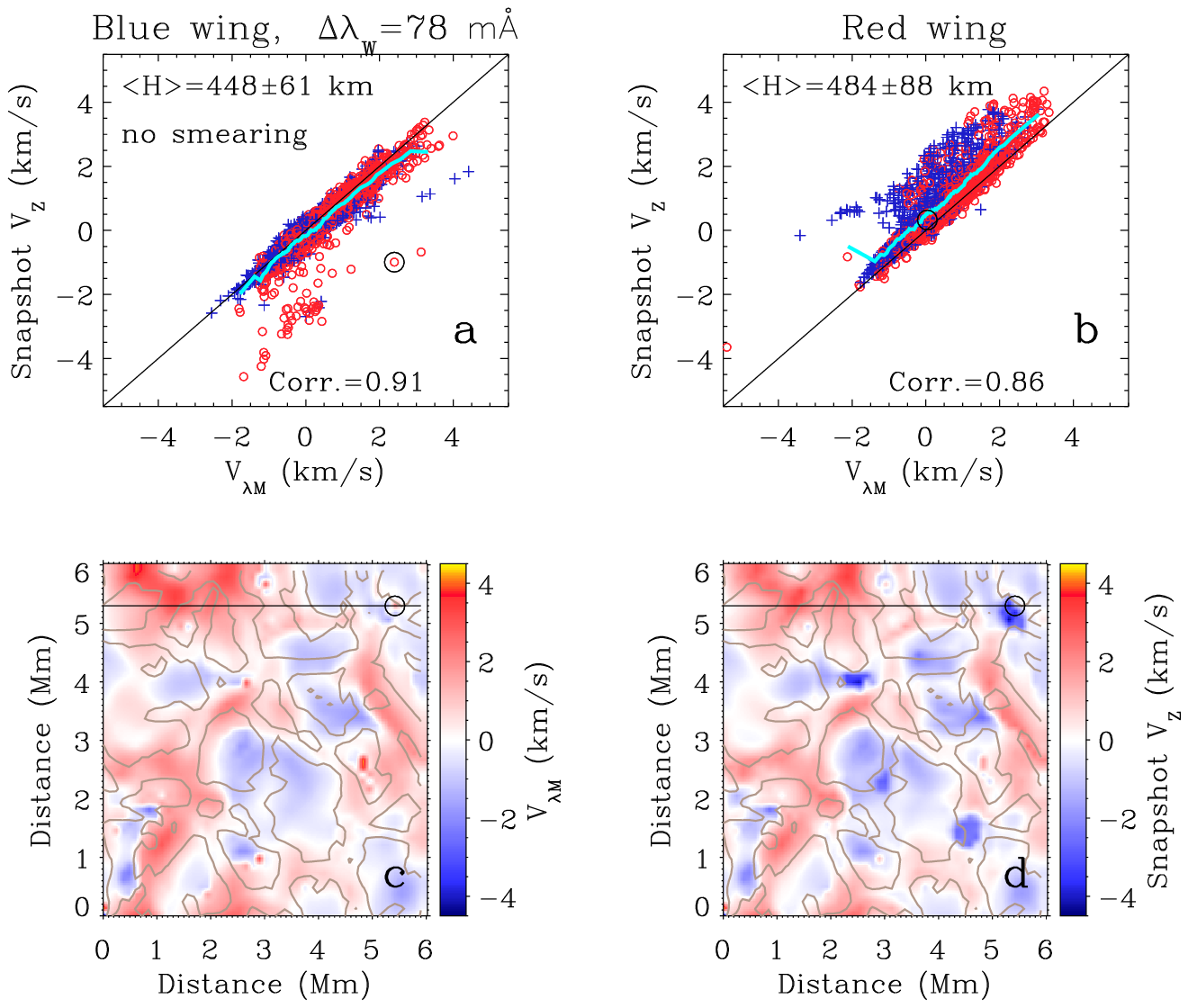}
\caption{Comparison between the $\lambda$-meter velocities
$V_{\lambda M}$ at $\Delta\lambda_W=$ 78 m\AA\ and snapshot
velocities $V_z$, measured at corresponding heights $H$ for the
case of no smearing. ({\it a}): Scatter plot of $V_{\lambda M}$
and $V_z$ in the blue wing. Black open circle marks the location
$x_i=5.5$ Mm, $y_i=5.4$ Mm. ({\it b}): Same for the red wing
velocities. ({\it c}): Map of the blue wing $V_{\lambda M}$. ({\it
d}): Map of the snapshot velocities $V_z$ corresponding to the
blue wing. The colour coding is the same as in
Figure~\ref{fig:parab_map}.} \label{fig:lambda2}
\end{figure*}

We slightly modified the standard $\lambda$-meter procedure
keeping in mind that  velocity field shifts the line opacity
coefficients of various atmospheric layers towards one wing of the
static line and away from the other. Such a shift  expands the
optical  line depth $\tau(\Delta {\lambda})$ on one side of
the line and compresses on the other side. As  a consequence, the
height of formation of the equal intensity points in the blue
$\Delta {\lambda}_b$ and red $\Delta {\lambda}_r$ wings belonging
the same spectral width $\Delta {\lambda}_W$  tends to be
different. Our modified $\lambda$-meter procedure is the
following. We applied the above set of spectral widths  both to
the spatially averaged  profile and to the individual profiles
over the snapshot. The spatially averaged profile corrected for
asymmetry due to granulation velocity field (see
Fig.~\ref{fig:hyper}, {\it bottom}) was considered as a reference
profile representing a stationary case. Since our study deals only
with the relative changes of the \BaII\ 4554 \AA\ line parameters
no correction for the asymmetry caused by the hyperfine structure
and isotopic shift was made. We derived the $\lambda$-meter
velocities $V_{\lambda M}$ in the blue wings by measuring the
Doppler shift of the blue intensity point belonging a certain
spectral width relatively the corresponding blue point of the
spatially averaged profile (same applies to the red wings).

The velocities $V_{\lambda M}$ obtained by this method were
compared to the corresponding snapshot velocities $V_z$. The $V_z$
values were specified separately for the blue $\Delta {\lambda}_b$
and red $\Delta {\lambda}_r$ wavelengths belonging the same line
width. These velocities were taken at heights where the optical
depth at the corresponding wavelength points was equal to unity (see
Sect. 2.2). The results of these calculation are displayed in
Figs~\ref{fig:lambda1}
to \ref{fig:lambda5-2}.

\subsection{$\lambda$-meter: original resolution}

Figure~\ref{fig:lambda1} shows the correlation coefficients
between the maps of $\lambda$-meter velocities $V_{\lambda M}$ and
the snapshot velocities $V_z$ taken at corresponding heights. It
gives several cases for different seeing conditions (value of the
parameter $R_0$) for the VTT-like data and also the case of no
smearing.
When the smearing is absent, the correlation coefficient for both
blue and red wing velocities is rather high. It reaches a maximum
value of $\sim 0.97$ around $\Delta {\lambda}_W  = 200$~m\AA,
being slightly smaller in the inner wing of the line.

The scatter plots and maps of the $V_{\lambda M}$ and $V_z$
velocities in the inner wing of the line at $\Delta {\lambda}_W  =
78$~m\AA\ are presented in Fig.~\ref{fig:lambda2} for the case of
no smearing. As follows from panel (a) of this figure, in the
inner blue wing $\lambda$-meter velocities agree essentially with
the true snapshot velocities at that height in most of the grid
points. The correlation coefficient between them is very high and
is close to 0.9. The comparison of the velocity maps in panels
(c) and (d) supports this conclusion. Nevertheless, there is a set
of grid points belonging mainly to intergranular lanes where the
fit is fairly bad.
The map of the true velocities $V_z$ in Fig.~\ref{fig:lambda2}d
reveals the presence of a number of intergranular lanes (as
defined by the continuum intensity) where the upflowing velocities
are observed at heights of formation of the blue wing intensity at
$\Delta {\lambda}_W  = 78$~m\AA. The $\lambda$-meter velocities do
not recover such upflowing points well, suggesting significantly
smaller absolute value of the velocities or even their sign
reversal. An example of such a location is marked by an open
circle in Figure~\ref{fig:lambda2}.

\begin{figure}
\centering
\includegraphics[width=7cm]{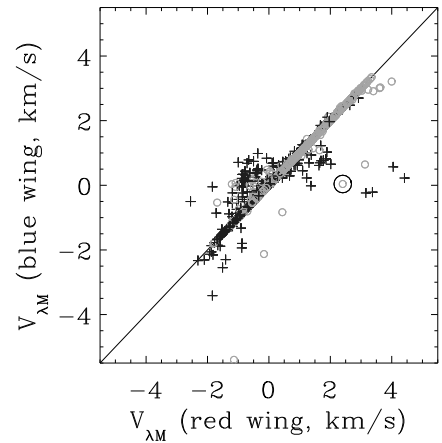}
\caption{Scatter plot of  the red and blue wing $\lambda$-meter
velocities at  $\Delta\lambda_W=$ 78 m\AA\ in the case of no
smearing. Black crosses are granular points and light circles are
intergranular points. the black open circle marks the point
corresponding the location $x_i=5.5$~Mm, $y_i=5.4$~Mm shown in
Fig.~\ref{fig:lambda2}. } \label{fig:lambda2d}
\end{figure}
%

Figure~\ref{fig:lambda2}b demonstrates that in the red inner wing
the match between $V_z$ and $V_{\lambda M}$ is appreciably worse.
The $\lambda$-meter velocities tend to be more redshifted.
According to Fig.~\ref{fig:lambda2d}, the blue and the red wing
$V_{\lambda M}$ are essentially the same (except for a few grid
points), whereas the corresponding $V_z$ are not. The reason for
that lies in the different formation heights of the blue and red
wing intensities at the same $\Delta {\lambda}_W$ (we discuss this
point in more details in Sect. 5.3). The velocities measured by
the $\lambda$-meter technique correspond better to the heights
where the blue-wing intensities are formed.

\begin{figure}
\centering
\includegraphics[width=9cm]{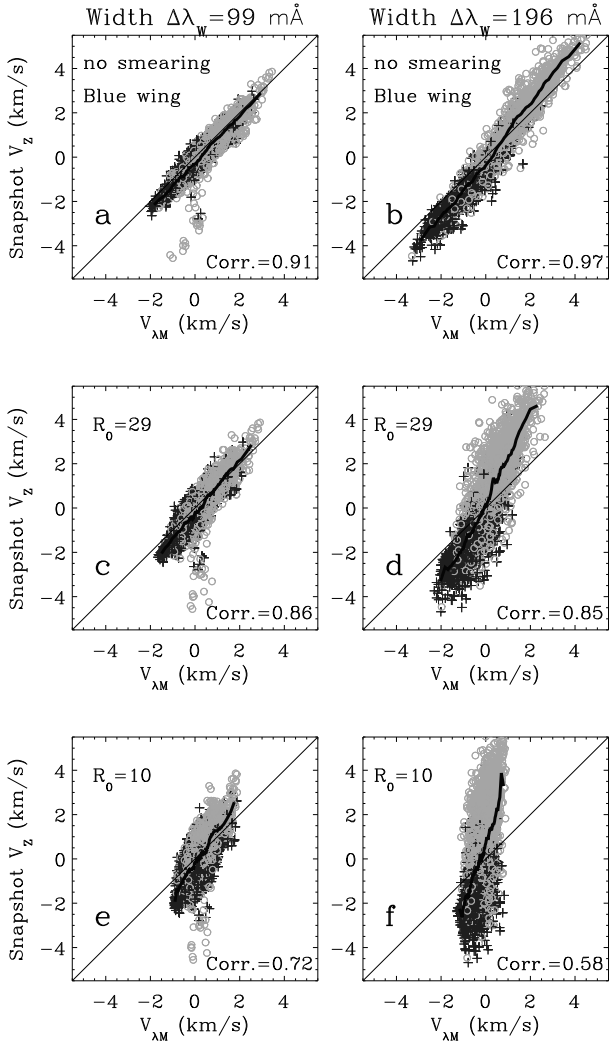}
\caption{Scatter plot of the $\lambda$-meter velocities
$V_{\lambda M}$  and the snapshot velocities $V_z$ in the wings of
the line at $\Delta {\lambda}_W = 99$~m\AA\ ({\it left panels})
and $\Delta {\lambda}_W = 196$~m\AA\ (right panels) for the
VTT-like data. ({\it a} and {\it b}): no smearing; ({\it c} and
{\it d}): Fried's parameter $R_0=29$; ({\it e} and {\it f}):
$R_0=10$. The gray scale coding is the same as in
Fig.~\ref{fig:lambda2d}.} \label{fig:lambda3h}
\end{figure}

Figure~\ref{fig:lambda3h}a, b gives two more examples of the
correlation between the $\lambda$-meter velocities and the
snapshot velocities in the outer wings of the line. The match is
typically rather good and the correlation is high, except that the
amplitudes of the $\lambda$-meter velocities are systematically
lower. The latter is easy to understand bearing in mind that the
method gives the average information over a certain height range,
thus leading to a decrease in the amplitude.

In summary, under perfect conditions, the $\lambda$-meter
technique allows us to obtain information about the LOS velocities
with a rather good precision over the whole photosphere, the
agreement being a little worse for the inner wings of the
line. The blue and the red wing $V_{\lambda M}$ velocities are
very close to each other.

\subsection{$\lambda$-meter: reduced resolution}

Apart from the case of the original resolution discussed above,
Fig.~\ref{fig:lambda1} contains the calculation of the correlation
coefficients between $V_z$ and $V_{\lambda M}$ for the VTT-like
data smeared to have a different spatial resolution by varying the
Fried parameter from 38 to 10 cm. As expected, the correlation
coefficient gets lower compared to the case of perfect seeing
conditions. Nevertheless, even for the medium spatial resolution
($R_0=10$ cm) the correlation for the inner wings ($\Delta
{\lambda}_W= 100$~m\AA) is rather high ($\sim 0.7$). For the
higher values of $R_0$ the correlation coefficient increases up to
0.8--0.9. Opposite  the case for no smearing, the correlation
coefficient decreases from the inner to the outer wings. The
velocity measured in the outer wings ($\Delta {\lambda}_W  >
120$~m\AA) is less precise.

The effects produced by the spatial smearing on the amplitudes of
the measured velocities $\Delta {\lambda}_W$ are shown in
Figure~\ref{fig:lambda3h}. The following points can be underlined from
this Figure:
\begin{itemize}
\item In the original data the maximum absolute values of the velocities are
higher above intergranular lanes than above granules. Spatial
smearing reduces both the absolute values of the velocities and
the asymmetry in the scatter plots. Nevertheless, the velocity
asymmetry is still present.
\item Even under excellent seeing conditions ($R_0=29$ cm) the outer
wing $\lambda$-meter velocities for $\Delta {\lambda}_W >
120$~m\AA\ are less reliable that the inner wing velocities.
\item The inner wing velocities ($\Delta {\lambda}_W <
120$~m\AA) still give a fairy good measure of the true velocities
for the Fried parameter ($R_0=10$~cm).
%
\item As the inner wings of the line are less sensitive to spatial
smearing, the \BaII\ 4554 \AA\ line can be useful to measure
velocities mainly in the upper photosphere.
\end{itemize}

\begin{figure*}
\centering
\includegraphics[width=14cm]{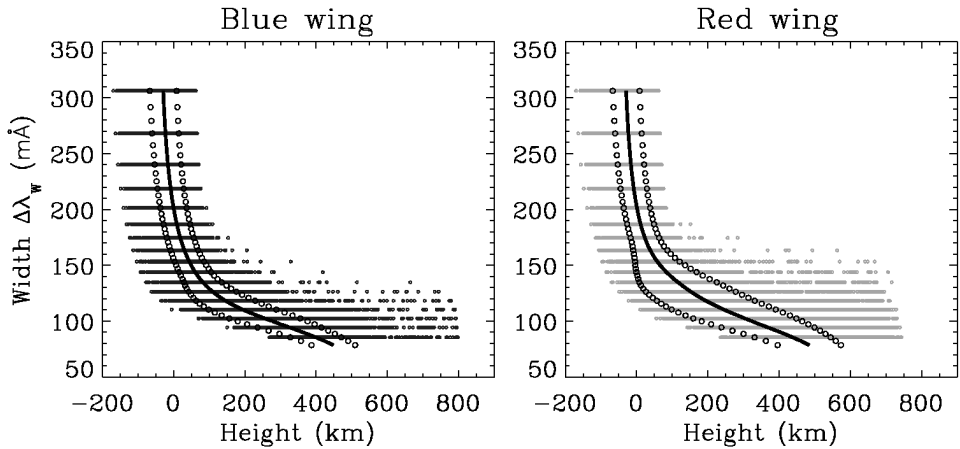}
\caption{Formation heights of the intensity along the line
profiles calculated for a set of spectral widths in the blue
(left) and red (right) wings for all grid points from the 3D
snapshot. Thick solid lines and open circles: mean formation
heights and their standard deviation, respectively.}
\label{fig:lambda5-1}
\end{figure*}

\subsection{Intensity formation heights along the line profile}

\begin{figure*}
\centering
\includegraphics[width=14cm]{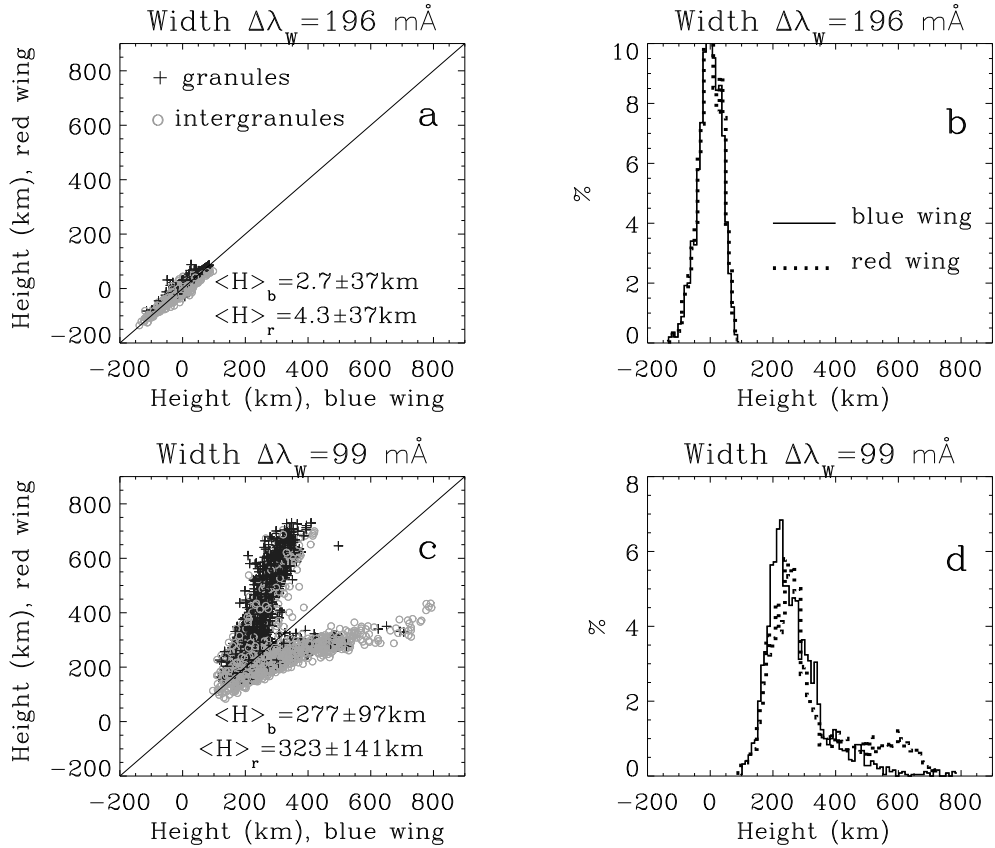}
\caption{ ({\it a, c}): Formation heights of the blue wing
intensities vs. formation heights of the red wing intensities at
two spectral widths: $\Delta {\lambda}_W = 196$~m\AA\ and $\Delta
{\lambda}_W = 99$~m\AA. Dark crosses - granular points; light
circles - intergranular points. ({\it b, d}): Histograms of the
formation heights of the blue wing intensities (solid line) and
red wing intensities (dotted line) for the same $\Delta
{\lambda}_W$. } \label{fig:lambda5-2}
\end{figure*}

The $\lambda$-meter technique only yields qualitative results as
long as it is not accompanied by the knowledge of heights where
the information on the velocity and intensity variations comes
from. In this section we give the results of the calculation of such
heights. We calculated the Eddington--Barbier formation heights
(see Sect. 2.2) for the intensity at each section of the \BaII\
4554 \AA\ line profile having a certain spectral width $\Delta
{\lambda}_W$. Below we use the notation $H_b$ for the blue wing
intensity formation heights calculated at $\Delta {\lambda}_b$
positions in the blue wing, and similarly for the red wing intensity
formation heights $H_r$, calculated at $\Delta {\lambda}_r$
positions in the red wing. Both $H_b$ and $H_r$ correspond to the
same $\Delta {\lambda}_W$. We repeated the calculation for the all
grid points of the 3D snapshot. The results of such calculations
are presented in Figures~\ref{fig:lambda5-1} and \ref{fig:lambda5-2}.
These figures give an answer to several important questions.

Firstly, how realistic is to assume that $H_b$ and $H_r$ belonging
to the same $\Delta {\lambda}_W$ are constant over the 3D
snapshot? The results shown in Fig.~\ref{fig:lambda5-1} suggest
that this assumption is far from reality. At each fixed $\Delta
{\lambda}_W$, both $H_b$ and $H_r$ vary in a rather wide range. In
outermost sections of the profile ($\Delta {\lambda}_W =
306$~m\AA) the heights vary between about $-175$ km and $+75$ km,
while in the inner section ($\Delta {\lambda}_W = 78$~m\AA) the
range of the variations is larger and lies between $+300$ km and
$+800$ km. Despite the large scatter, the mean formation heights
of the each section of the profile $\left < H \right >$ have a
well pronounced dependence on $\Delta {\lambda}_W$ (thick solid
curves in Fig.~\ref{fig:lambda5-1}). This  makes it possible to assign
in, a certain way, a height dependence to the velocity
measurements by $\lambda$-meter technique. This can be done by
ascribing the response of each $\Delta {\lambda}_W$ section of the
spectral line to the mean height $\left < H \right >$.

Secondly, is it correct to assume that the $H_b$ and $H_r$
formation heights belonging to the same $\Delta {\lambda}_W$ are
equal? In order to answer this question, we show in
Fig.~\ref{fig:lambda5-2} two representative cases for $H_b$ and
$H_r$ calculated at two positions in the inner ($\Delta
{\lambda}_W=99$ m\AA) and the outer ($\Delta {\lambda}_W=196$
m\AA) wings of the line. The $\Delta {\lambda}_W = 196$~m\AA\
corresponds to the section of the \BaII\ 4554 \AA\ where the line
opacity profile is shallow. So one can expect the difference in
heights of formation $H_b$ and $H_r$ caused by the Doppler shift
of the opacity profile to be small. The
width $\Delta {\lambda}_W  = 99$~m\AA\ belongs to the inner
section of the \BaII\ 4554 \AA\ characterized by a steeper part of
the line opacity  profile. It is expected that the same Doppler
shifts produce a more important difference in the optical depths
on opposite sides of the \BaII\ 4554 \AA\ line and, hence, a
somewhat greater difference between the $H_b$ and $H_r$ heights.
What is the magnitude of this difference?

As follows from Fig.~\ref{fig:lambda5-2}a,  in the outer wings of
the \BaII\ 4554 \AA\ line the blue $H_b$ and red $H_r$ heights are
indeed very close to each other, both varying over the 3D
snapshot. The histograms displayed in Fig.~\ref{fig:lambda5-2}b
demonstrate that the distributions of $H_b$  and  $H_r$ are
similar and have a sharp cut off. It means that outer wings are
formed in a narrow atmospheric layer with mean heights $\left <
H_b \right >$ and $\left < H_r \right >$ of the blue and red wings
being very close to each other.

Figure~\ref{fig:lambda5-2}c shows a similar calculation for the
inner sections of the \BaII\ 4554 \AA\ line. The Doppler shift of
the line opacity profile leads to a complex behaviour of the $H_b$
and $H_b$ heights. Two groups of points can be distinguished in
the scatter plot of $H_r$ vs. $H_b$ calculated over the snapshot.
For the first group (symbols above diagonal), the $H_b$ heights
are concentrated in a rather narrow band around $+200$ km while the
$H_r$ heights extend for several hundreds of kilometres from nearly
$+100$ km to $+800$ km. This group of points belong largely to
the profiles coming from granular regions. For the second group of
 points (symbols below the diagonal) the behaviour is the
opposite. This group originates mostly in intergranular lanes.
Only a relatively small number of the profiles have approximately
the same $H_b$ and $H_r$ heights (symbols along the diagonal).

The histograms of the $H_b$ and $H_r$ heights for the inner
sections of the \BaII\ 4554 \AA\ line (Fig.~\ref{fig:lambda5-2}d)
are not symmetric. The histograms have a maximum at heights around
$+200$ km and a sharp cut-off below this height. There is a long
tail toward 600--800 km. This tail is more pronounced for the red
wing heights $H_r$. Such an asymmetry suggests that the
information in the inner wing comes from two distinct layers. In
most of the grid points the intensity at $\Delta
{\lambda}_W=99$ m\AA\ is formed around $+300$ km, whereas in a
smaller but still appreciable number of points the intensity comes
from higher layers between $+400$ and $+800$ km. On the whole, the
mean formation height of the inner red wing is  larger than that
of the inner blue wing.

\section{Conclusions}

In this paper we have analysed the range of validity of the two
Doppler diagnostic techniques using the \BaII\ 4554 \AA\ line,
\ie\ the 5-point Dopplergram technique \citep{sutt:etal:2001} and
the $\lambda$-meter technique \citep{steb:good:1987}. We have
performed NLTE radiative transfer calculations of the \BaII\ 4554
\AA\ intensity profiles in the 3D snapshot of hydrodynamical
simulations of solar convection \citep{asp:etal:2000}, neglecting
the effects of horizontal radiative transfer, but considering a
realistic barium atomic model and taking into account the
hyperfine structure and isotopic shift. The original resolution
profiles were smeared to reproduce the DOT-like and VTT-like data
and study the effects of the limited resolution into the
reliability of the results produced by the two Doppler diagnostic
techniques.

Our results for the 5-point Dopplergram technique can be
summarized as follows:

\begin{itemize}
\item The NLTE simulations using the 3D hydrodynamical model support
the opinion that the speckle-reconstructed Dopplergram velocities
obtained from the DOT-like data give appropriate representation of
the solar photospheric velocity field.
\item{The information on the velocities obtained from such
Dopplergrams comes from a thin atmospheric layer located at
heights around 300 km.  The speckle-reconstructed Dopplergram
velocity maps reproduce in many details the ``true'' velocity
structures existing in this layer.}
\item{The Dopplergram technique can overestimate the velocities
in  intergranular areas with strong downflows. This excess appears
in the speckle-reconstructed maps as localized points with
enhanced brightness. The interpretation of such bright points in
terms of magnetic fields has to be carried out with caution. At
least some of them may be an artefact caused by the Dopplergram
technique itself.}
\end{itemize}

Our conclusions regarding the $\lambda$-meter technique are the
following:

\begin{itemize}
\item{Under perfect seeing conditions the $\lambda$-meter
technique allows us to obtain information about the LOS
velocities throughout the photosphere with  rather good
precision. Only in the upper photosphere is the particular velocity
structure with upflowing points in intergranular lanes not
always well reproduced.}
\item{The velocities measured by the $\lambda$-meter technique
correspond better to the ``true'' snapshot velocities taken at
heights of formation of the blue line wing, rather that the red
wing.}
\item{Even for rather medium seeing conditions the inner wings
of the \BaII\ 4554 \AA\ line give reliable information about the
velocity field in the upper photosphere. The results from the
outer wings are less reliable.}
\item{The mean formation heights of each section of
the \BaII\ 4554 \AA\ line profile have a well-pronounced
dependence on the spectral width of the section. This gives the
possibility to assign height dependence to the velocity
measurements by the $\lambda$-meter technique. The non-thermal
motions can be reliably measured with the $\lambda$-meter
technique applied to the \BaII\ 4554 \AA\ line throughout the
photosphere up to the temperature minimum.}
\end{itemize}

\begin{acknowledgements}
This work was partially supported by the Spanish Ministerio de
Educaci{\'o}n y Ciencia (projects AYA2007-63881 and
AYA2007-66502), and by the National Academy of Sciences of Ukraine
(project 1.4.6/7-233B).
\end{acknowledgements}


\begin{appendix}

\section{NLTE modeling}
\label{section:NLTE}

\begin{figure}
\centering
\includegraphics[width=8.5cm]{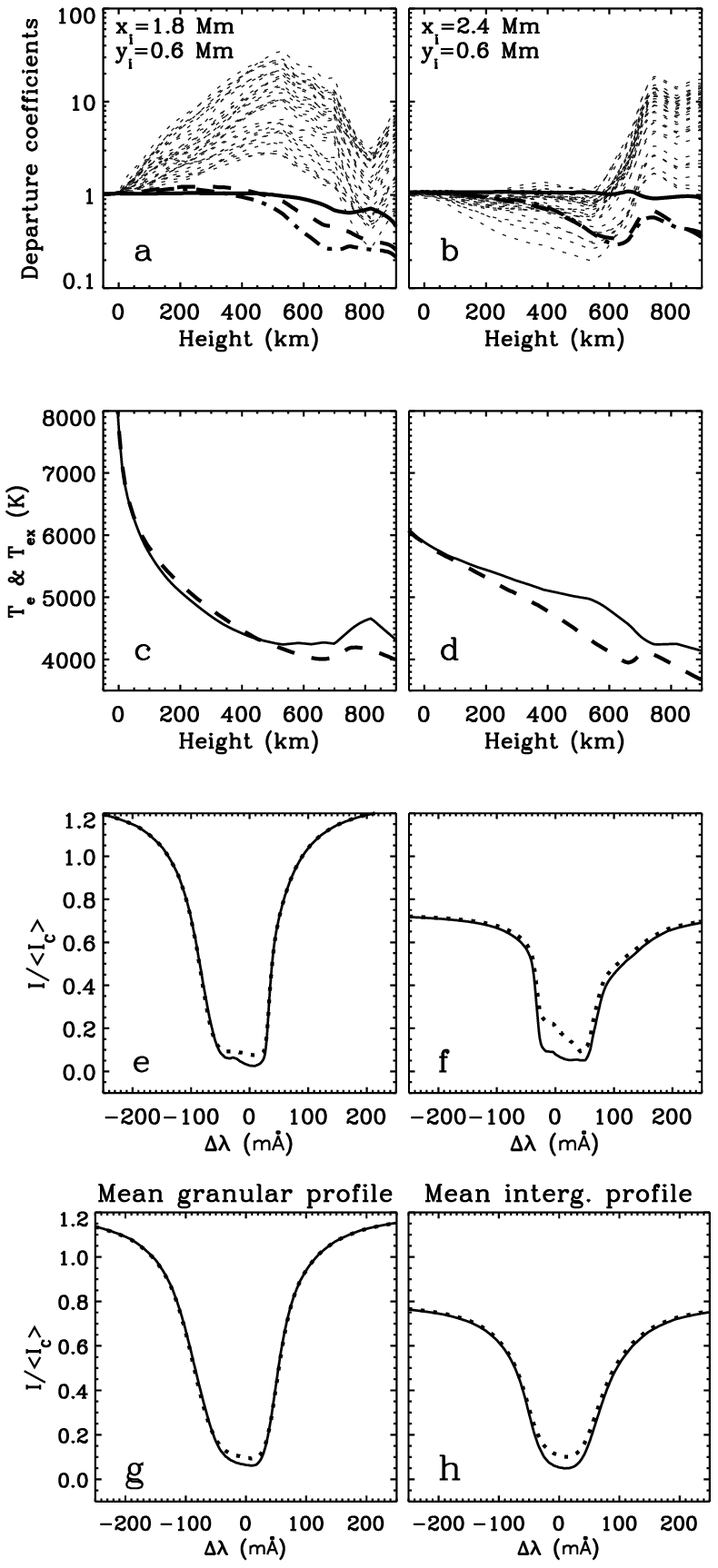}
\caption{({\it a, b}): Departure coefficients $\beta$ of the
\BaII\ atomic levels  vs. height in a representative granular
({\it left}) and intergranular ({\it right}) models.  Thick solid
and thick dash lines: the $\beta$-coefficients for the ground
$6s\, \rm{{^2}S}_{1/2}$ and upper $6p\, \rm{{^2}P}_{3/2}$ levels
of the \BaII\ 4554 \AA\ line, respectively. Dash-dotted lines: the
$\beta$-coefficients for the upper level of the \BaII\ 4934 \AA\
resonance line. Dotted lines: the $\beta$-coefficients for the
\BaII\ levels with excitation potentials  above 5 eV. ({\it c,
d}): The line source function $S_L$ of the \BaII\ 4554 \AA\ line
(dash) and Planck function $B$ (solid) in units of the temperature
vs. height for the same models. ({\it e, f}): The \BaII\ 4554 \AA\
line profiles for the representative granule and itergranule.
Solid and dotted lines: NLTE and LTE, respectively. ({\it g, h}):
Spatially averaged granular and intergranular profiles. }
\label{fig:beta}
\end{figure}


\begin{figure}
\centering
\includegraphics[width=8.5cm]{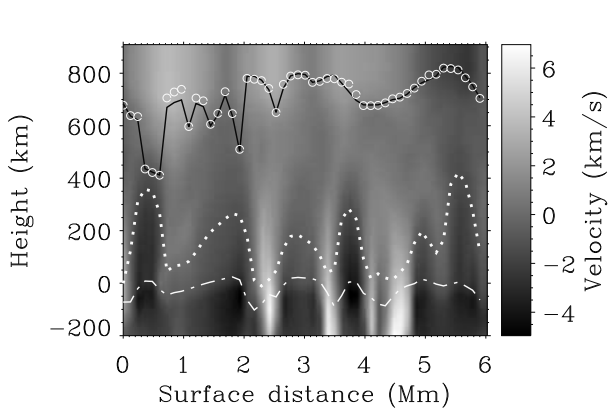}
\caption{The  NLTE (solid line) and LTE (open circles) heights of
formation of the core of the \BaII\ 4554 \AA\ line along the slice
of the snapshot $y_i=0.6$ Mm. Dash-dotted line: continuum height
of formation at 4554 \AA. Dotted line: height of formation of the
line wing for the wavelength position $\Delta \lambda=-76.5$~m\AA.
The background image is the snapshot vertical velocity $V_z$.
Negative (upflow) velocities $V_Z$ correspond to granules (dark),
while positive (downflow) velocities to intergranules (light).}
\label{fig:NLTE_height}
\end{figure}

\begin{figure}
\centering
\includegraphics[width=9cm]{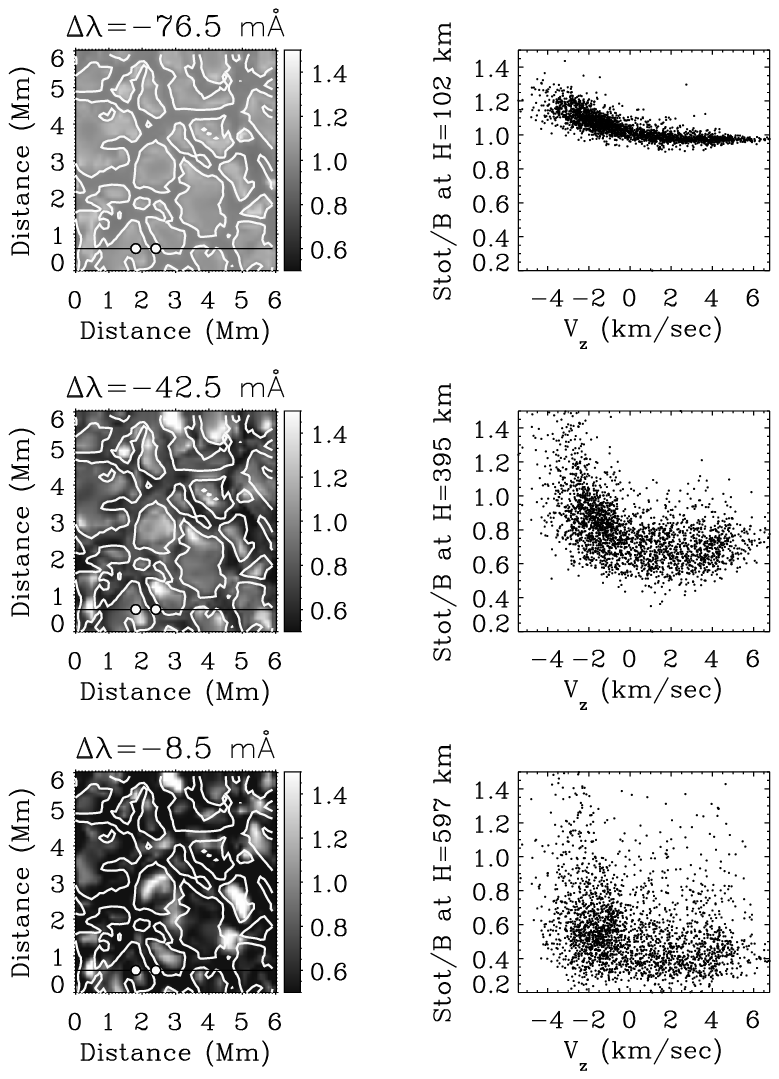}
\caption{The total source function $S_{tot}(\Delta \lambda)/B$
measured in the Planck function units   for three wavelength
points $\Delta \lambda=-76.5; -42.5; -8.5$~m\AA\ situated  in the
blue wing of the Ba\,{\sc ii} 4554 \AA\ line profile. {\it Left}:
Maps of the $S_{tot}(\Delta\lambda)/B$ at the mean intensity
formation heights at these wavelengths. Horizontal line
corresponds to the slice $y_i=0.6$ Mm and the filled circles to
the surface positions $x_i=1.8$ Mm (granule), $x_i=2.4$ Mm
(intergranule). {\it Right}: Scatter plots of $S_{tot}(\Delta
\lambda)/B$ and velocities $V_z$ for the same wavelengths. The
velocities are  taken at heights of formation of continuum
intensity. } \label{fig:stot}
\end{figure}


Figure~\ref{fig:beta} (panels {a} to {f}) shows the population
departure coefficients, the \BaII\ $\lambda$4554 \AA\ line source
functions and line profiles for two spatial grid points of the
3D snapshot representing  the typical granular and intergranular
models. We use these models to illustrate the difference in the
NLTE results for granules and intergranules. The population
departure coefficients are defined as $\beta =n_{\rm NLTE}/n_{\rm
LTE}$ where $ n_{\rm NLTE}$ and $n_{\rm LTE}$  are the NLTE and
LTE atomic level populations, respectively. The Complex behaviour
of the $\beta$-coefficients shown in Fig.~\ref{fig:beta}a, b is a
result of the interaction of several NLTE mechanisms described in
detail by \citet{carlsson:etal:1992, bruls:etal:1992,
Shchukina:Trujillo:2001}. Here we just point out that for the
barium atom the most important of them are ultraviolet line
pumping, ultraviolet overionization, resonance line scattering and
photon losses.
The resonance line scattering and photon losses manifest
themselves as a divergence of the lower $6s\, \rm{{^2}S}_{1/2}$
and upper $6p\,\rm{{^2}P}_{1/2}$, $6p\, \rm{{^2}P}_{3/2}$ levels
of the \BaII\ resonance lines. This divergence results from the
surface losses near the layer where the optical depth is equal to
unity. The losses propagate by scattering to far below that layer.
Interestingly, for the integranule the divergence of the
$\beta$-coefficients arises in the innermost layers. This happens
because the photon losses occur mainly through the line wings of
the \BaII\ 4554 \AA\ line. As follows from
Fig.~\ref{fig:NLTE_height} the line wings in integranules are
formed considerably deeper than in granules. Such a difference in
the formation heights is a result of the Doppler shift of the line
opacity coefficient caused by the velocity field.  As a
consequence, in the intergranular model (see
Fig.~\ref{fig:beta},\,{a, b}) the divergence starts already in the
lower photosphere while in the granular model it happens only in
upper photosphere at heights around 400 km.

Another important conclusion that follows from
Fig.~\ref{fig:NLTE_height} concerns the height of formation of the
\BaII\ 4554 \AA\ line. The lower departure coefficient $\beta_l$
is close to unity. So the scaling of the line opacity with this
coefficient cannot lead to an appreciable difference between the
NLTE and LTE heights of formation of this line.

The excess of \BaII\  ions at the levels with excitation
potentials  above 5 eV  visible in the granule model is produced
by the pumping via the  ultraviolet \BaII\   lines starting at
$6s\, \rm{{^2}S}$, $5d\, \rm{{^2}D}$, $6p\, \rm{{^2}P}^o$ levels.
For the intergranule model the overpopulation arises only in the
uppermost layers. Such behaviour of the $\beta$-coefficients
corresponds to the temperature stratification of the models. The
overpopulation of the high excitation levels of \BaII\ in granules
occurs because here the excitation temperature of the ultraviolet
pumping radiation field appreciably exceeds the electron
temperature. In integranules such superthermal radiation, and
hence  the level overpopulation,  is present  only above the
temperature minimum region. In addition, in the intergranular
photospheric layers the photon losses in the ultraviolet lines are
more pronounced than in granules.

The ${\beta}_u / {\beta}_l$ ratio of the upper and lower level
departure coefficients of  the \BaII\ 4554 \AA\ line sets the
departure  of its line source function $S_L$ from the Planck
function $B$. Figure~\ref{fig:beta}\,({c, d}) shows that this
departure (reflecting the corresponding departure coefficient
divergence in the upper panels of this Figure) is larger in the
intergranular than in the granular model.

Figure~\ref{fig:stot} demonstrates that such  behaviour is typical
also for the total source function $S_{\rm {tot}}$ at the
wavelengths corresponding to the inner wings ($\Delta \lambda  <
76.5$~m\AA). On average, in intergranular regions it drops below
the Planck function while in granules the effect is less
pronounced. Moreover, in granular areas with strong upflows the
total source function can exceed the Planck function. This excess
can be understood if one takes into account that the resonance
source line function is described by the two-level approximation,
i.e. it approximately equals mean intensity $J$. In the regions
with small photon losses (like granules) the $J> B$, hence,
$S_{L}$ and $S_{\rm {tot}}$ have to be greater than $B$ as well.

Figure~\ref{fig:beta}\,({e, f}) show the NLTE and LTE disc-centre
line profiles for the individual  granular and integranular
models. The profiles displayed in Fig.~\ref{fig:beta}\,(g ) result
from averaging of the emergent intensities corresponding only to
the granular models. Averaged intergranular profiles are shown in
Figure~\ref{fig:beta}\,(h). These two bottom panels quantify the
statistical effect produced by the deviation from the LTE in two such
types of the atmospheric models. The main conclusions that may
be drawn from the results presented in
Fig.~\ref{fig:beta},~\ref{fig:NLTE_height},~\ref{fig:stot} are the
following:
\begin{itemize}
\item
{The source function deficit, as compared to the LTE assumption,
is the main mechanism that controls the formation of  the Ba\,{\sc
ii} $\lambda$4554 \AA\   line. The line opacity deficit is small
and, hence, unimportant.}
\item
{The divergence between $S_{tot}$ and  $B$ changes the shape of
the individual profiles, particularly  the intergranular ones.}
\item
{On average, the deviations from the LTE lead to deepening (i.e.
strengthening) of  the spatially averaged \BaII\ $\lambda$4554
\AA\ line profiles. The NLTE effects are most pronounced around
the line core and are generally more important in the
intergranular regions than in the granular ones. The mean difference
between the NLTE and LTE line core residual intensities does not
exceed 5\% for granules and 10\% for intergranules.}
\item
{Towards the wings, the LTE becomes a valid description for the
\BaII\ $\lambda$4554 \AA\   line profile.}
\end{itemize}

\end{appendix}

\end{document}